\documentclass[11pt]{article}
\usepackage{amssymb,latexsym, amsmath}

\def\setmarsing{
\oddsidemargin-0.01in
\textwidth6.4in
\textheight8.5in}
\setmarsing

\makeatletter
\@addtoreset{figure}{section}
\def\thefigure{\thesection.\@arabic\c@figure}
\def\fps@figure{h,t}
\@addtoreset{table}{bsection}

\def\thetable{\thesection.\@arabic\c@table}
\def\fps@table{h, t}
\@addtoreset{equation}{section}

\makeatother

\newtheorem{theorem}{Theorem}[section]

\newtheorem{corollary}[theorem]{Corollary}

\begin{document}

\title{K\'arm\'an--Howarth Theorem for the \\
 Lagrangian averaged Navier-Stokes alpha model}
\author{
Darryl D. Holm
\\Theoretical Division and Center for Nonlinear Studies
\\Los Alamos National
Laboratory, MS B284
\\ Los Alamos, NM 87545
\\ {\footnotesize email: dholm@lanl.gov}
}
\date{May 16, 2001}
\maketitle



\begin{abstract}
The K\'arm\'an--Howarth theorem is derived for the Lagrangian averaged
Navier-Stokes alpha (LANS$-\alpha$) model of turbulence. Thus, the
LANS$-\alpha$ model's preservation of the fundamental transport structure
of the Navier-Stokes equations also includes preservation of the transport
relations for the velocity autocorrelation functions. This result implies
that the alpha-filtering in the LANS$-\alpha$ model of turbulence does
not suppress the intermittency of its solutions at separation distances
large compared to alpha.
\end{abstract}

\section{Introduction}

\subsection*{K\'arm\'an-Howarth (KH) theorem}

The dynamics of the two-point velocity autocorrelation functions in
turbulence contains information complementary to the spectral
description. These autocorrelation functions directly connect the
concept of scale with the result of an actual flow measurement. However,
the autocorrelation functions yield no information about the energy, or
enstrophy, that is contained in their interval of separation.
Instead, the third-order correlation functions do provide information
about the {\it fluxes} of energy and enstrophy as functions of separation
distance.  Thus, the spectral and autocorrelation dynamics give
complementary information about the same phenomenon. The KH theorem
relates the time derivative of the two-point velocity autocorrelation
functions to the divergences of the third-order correlation functions.
Perhaps not surprisingly, because of its fundamental importance, the line
of investigation that began in 1938 with the KH theorem is still being
actively pursued.

The invariant theory of isotropic turbulence was introduced by
von K\'arm\'an \& Howarth (1938) and refined by Robertson (1940), who
reviewed the K\'arm\'an--Howarth (KH) theorem in the light of classical
tensor invariant theory. The physical importance of the KH
theorem in turbulence modeling is undeniable. According to Monin
\& Yaglom (1975) (vol II, page 122) the KH theorem's dynamical equation
for the two-point autocorrelation function of the fluid velocity, ``plays
a basic part in all subsequent studies in the theory of isotropic
turbulence.'' A homogeneous (but not necessarily isotropic) version of
this theorem was introduced by Monin (1959). This non-isotropic variant
of the KH theorem was discussed further by Monin \& Yaglom (1975) (vol
II, page 403), but there was a gap in the proof. Correct proofs were
given independently by Frisch (1995) and Lindborg (1996). These proofs
are reviewed in Hill (1997), who concentrates on the logical steps
needed to eliminate pressure-velocity correlations in the KH
theorem without assuming isotropy. 

The isotropic K\'arm\'an--Howarth theorem was extended to
magnetohydrodynamics (MHD) in Chandrasekhar (1951) and was
recently analyzed further using Els\"asser variables by Politano \&
Pouquet (1998). Extensions of the KH theorem and its corollary --
the ``$-4/5$ law'' of Kolmogorov (1941b) -- to include passively
advected scalars such heat and chemical concentration (or buoyancy) were
given by Yaglom (1949) and were later refined by Antonia, Ould-Rouis,
Anselmet \& Zhu (1997), who also compared with Yaglom's results for the
case of stratified fluids. The effect of fluid helicity on the KH
theorem was discovered in Chkhetiani (1997) and analyzed in terms of
structure functions in Gomez, Politano \& Pouquet (1998). Arad, L'vov \&
Procaccia (1997) extended the fundamental results in Robertson (1940) by
considering projections of the fluid velocity autocorrelation dynamics
onto other irreducible representations of the $SO(3)$ symmetry group,
besides the scalars under rotation. Modern computational resources are
frequently applied, such as in Fukayama et al. (2000) and
references therein, for numerically studying intermittency in turbulence,
which is represented by the anomalous scaling behavior of its higher
order structure functions, relative to those in the KH theorem.

\subsection*{Navier-Stokes alpha model}

Recently, a modification of the Navier-Stokes (NS) equations known as   
the Navier-Stokes alpha model was introduced and its solution
behavior was compared with experimental data for turbulent flow in pipes
and channels in Chen et al. (1998, 1999a, 1999b). These modified equations
essentially filter the fluid motion that occurs below a certain length
scale (denoted as alpha), which is a parameter in the model. Amongst
other differences from traditional turbulence modeling approaches, this
alpha-filtering approach differs from the large eddy simulation (LES)
approach by preserving the basic transport structure of the exact NS
equations.  Direct numerical simulations (DNS) of this model for forced
homogeneous turbulence were performed in Chen et al. (1999c). Homogeneous
turbulence decay was also simulated numerically using this model in Mohseni
et al. (2000). Both these simulation studies showed the NS$-\alpha$ model
to be considerably more computable than the exact NS equations, while
preserving essentially the same solution behavior as NS at length scales
larger than alpha. A review of the basic properties of the Navier-Stokes
alpha model and its early development is given in Foias, Holm and Titi
(2001a). See also Foias, Holm and Titi (2001b) and Marsden and Shkoller
(2001) for recent analytical and geometrical results for this model. 

The alpha model's original derivation (without viscous dissipation) in
Holm, Marsden and Ratiu (1998a, 1998b) was motivated by seeking a 
generalization to $n$ dimensions of the integrable one-dimensional
shallow water equation due to Camassa and Holm (1993).  This derivation
was based on the Lie algebraic and geometric properties of Hamilton's
principle for fluid dynamics in the Eulerian description. However,
motivated by its later derivation as a Lagrangian averaged fluid model
in Holm (1999), Holm (2001), and by a related  derivation in Marsden and
Shkoller (2001), we shall henceforth refer to this model as the
Lagrangian-averaged Navier-Stokes alpha model, or the LANS$-\alpha$
model.

Although several analytical and geometrical results are known for the
LANS$-\alpha$ model, many open questions still remain. An outstanding 
question concerns the effect of alpha on intermittency in turbulence. One
supposes that filtering at scales smaller than alpha would suppress
intermittency on those scales. However, what is the effect of
alpha-filtering at small scales on intermittency in the larger scales?
Proving the KH theorem for the LANS$-\alpha$ model is a {\it required}
first step toward addressing this outstanding question. Remarkably, the
effects of the alpha-filtering are confined to separations of order
$O(\alpha)$ or smaller, in the analogs of the KH theorem and in
Kolmogorov's energy balance law proved here for the LANS$-\alpha$ model.

The KH theorem for the LANS$-\alpha$ model proved here provides an
exact result that demonstrates how the introduction of the length scale
alpha affects the dynamics of the velocity autocorrelations as a function
of separation between two points fixed in the domain of an isotropic
LANS$-\alpha$ flow. These effects turn out to be negligible at
separations that are large compared to the filtering scale alpha
($r\gg\alpha$). To the extent that the same result persists in the
higher correlation functions and structure functions for the LANS$-\alpha$
model, one may expect the intermittency of the alpha model to be the same
as that for Navier-Stokes at large separations ($r\gg\alpha$).

\subsection*{The LANS$-\alpha$ model equations}

The motion equation for the unforced LANS$-\alpha$ model is given by
\begin{eqnarray*} 
\partial_t \mathbf{v}
+
\mathbf{u}\cdot\nabla\mathbf{v}
+
v_j\nabla u^j
+
\nabla
\Big(\,p - \frac{1}{2}|\mathbf{u}|^2  
- \frac{\alpha^2}{2}|\nabla\mathbf{u}|^2  \,\Big)
=
\nu\Delta\mathbf{v}
\,,\\
\hbox{where}\quad
\mathbf{v}
\equiv
\mathbf{u}-\alpha^2\Delta\mathbf{u}
\,,\quad
\nabla\cdot\mathbf{u}=0
\,.
\end{eqnarray*} 
The left side of this motion equation preserves the transport structure of
ideal fluids, especially the Kelvin-Noether theorem, see Holm, Marsden
and Ratiu (1998a, 1998b). Here, the specific momentum $\mathbf{v}$ is
related to the fluid velocity
$\mathbf{u}$, via the Helmholtz operator $1-\alpha^2\Delta$ which
contains the length scale $\alpha$. Although derived from a different
approach, these equations are formally similar to the equations for the
motion of a 2$^{nd}$ grade fluid, but with a different dissipation
operator. Following Camassa \& Holm (1993) for the original
one-dimensional version of this equation, the motion equation for the
alpha model may also be expressed equivalently in a way that emphasizes
its nonlocality, as 
\begin{eqnarray} \label{LANS-alpha-motion-eqn}
(1-\alpha^2\Delta)
\Big(\,
\partial_t \mathbf{u}
+
\mathbf{u}\cdot\nabla\mathbf{u}
-
\nu\Delta\mathbf{u}
\,\Big)
+
\nabla p 
=
-\,{\rm div}\tau
\,,\\
\hbox{where}\quad
\tau
\equiv
\nabla\mathbf{u}\cdot\nabla\mathbf{u}
+
\nabla\mathbf{u}\cdot\nabla\mathbf{u}^T
-
\nabla\mathbf{u}^T\cdot\nabla\mathbf{u}
\,,\quad
\nabla\cdot\mathbf{u}=0
\,.\nonumber
\end{eqnarray} 
%
Similar forms of the motion equation for a 2$^{nd}$ grade fluid
also appear in Dunn \& Fosdick (1974), modulo a crucial 
difference in dissipation.%
\footnote{The formal similarity of the LANS$-\alpha$ model with the
2$^{nd}$ grade fluid equation (8.1) on page 241 of Dunn \& Fosdick (1974)
resonates with a remark made already in von K\'arm\'an \& Howarth (1938)
page 197, ``The expression $\dots$ for the correlation tensor is exactly
of the same form as that for the stress tensor for a continuous medium
when there is spherical symmetry.'' Mathematically, the crucial
difference  between the LANS$-\alpha$, and 2$^{nd}$ grade fluid models 
lies in their different dissipation operators. In the motion equation 
(\ref{LANS-alpha-motion-eqn}) for the fluid velocity, $\mathbf{u}$, the
LANS$-\alpha$ model has diffusive viscous  dissipation
$-\,\nu\Delta\mathbf{u}$, introduced in Chen et al. (1998, 1999a, 1999b)
for their comparisons with turbulent flow data, while the 2$^{nd}$ grade
fluid has only damping. That is,
$-\,\nu\Delta\mathbf{u}$ in equation (\ref{LANS-alpha-motion-eqn}) is
replaced by $\mu\mathbf{u}$ in the motion equation for 2$^{nd}$ grade
fluids, for a positive constant
$\mu$.  Physically, there are also significant differences in the
interpretations of both the solutions and the parameters that appear in
the equations for the two models. }
Upon inverting the Helmholtz operator in
three dimensional Cartesian components, this becomes
\begin{eqnarray*} 
&&\hspace{-1cm}\partial_t u_i
+
\partial_k\big(\,u_i u^k + \tilde{p}\,\delta_i^k \,\big)
-
\nu\Delta{u}_i
=
-\,\alpha^2\,\partial_k\, \Big(g_\alpha*\tau_i^k\Big)
\,,
\
\hbox{where}\,i,j,k\in1,2,3,
\\&&\hspace{-1cm}
g_\alpha*\tau_i^k
\equiv
\int
g_\alpha(|\mathbf{x}^{\,\prime}-\mathbf{x}|)
\,\tau_i^k(\mathbf{x}^{\,\prime})\,d^3{x}^{\,\prime}
\,,\
\tau_i^k
\equiv
\Big(\,
u_{i,j}u^{j,k}
+
u_{i,j}u^{k,j}
-
u_{j,i}u^{j,k}
\,\Big)
\,.
\end{eqnarray*} 
Here $(1-\alpha^2\Delta)\tilde{p}=p$ and $g_\alpha$ is the Green's
function for the Helmholtz operator in the flow domain with 
boundary conditions compatible with {\it satisfying NS on the boundary}.
The free space Green's function for the Helmholtz operator in three
dimensions is $g_\alpha(r)=r^{-1}\exp{-(r/\alpha)}$,  which is the
well-known Yukawa potential for $r=|\mathbf{x}^{\,\prime}-\mathbf{x}|$.
This free space Green's function should apply to the extent that the
turbulence is isotropic (so that $\alpha$ can be taken as constant) and
away from boundaries at distances greater than order $O(\alpha)$. 

This form of the LANS$-\alpha$ equation demonstrates that
the nonlinear stresses proportional to $\alpha^2$ are filtered by the
Green's function $g_\alpha(r)$ of the Helmholtz operator, whose filter
width is of order $O(\alpha)$. Given $\alpha$, the term
$g_\alpha*\tau_i^k$ applies at separations $r$ such that $\alpha/r\le1$ and
is negligible for $\alpha/r\ll1$. For those larger separations, the
LANS$-\alpha$ equation reverts to NS. The alpha-filtering is a
regularization. With it, the solutions of LANS$-\alpha$ are well-posed and
have a {\it finite dimensional global attractor} under $L_2-$bounded
forcing, as shown in Foias, Holm and Titi (2001b). See also  Marsden and
Shkoller (2001) for a proof of their well-posedness in the presence of
boundaries. Such well-posedness results are not known to hold for the NS
equations. However, solutions of the LANS$-\alpha$ equations do converge to
solutions of the NS equations uniformly as $\alpha\to0$ for any positive
viscosity, as also shown in Foias, Holm and Titi (2001b). 

\subsection*{The equations governing the correlations}

In deriving the correlation dynamics, one may regard
$(\,g_\alpha*{\tau^{\,\prime}}_j^k\,)$ as a subgrid scale (actually,
sub$-\alpha$ scale) stress tensor arising from the alpha-filtering
procedure represented by convolution with the Green's function 
$g_\alpha(r)$, which decreases exponentially in separation with
scale length $\alpha$. See Holm (1999), Holm (2001) and Marsden and
Shkoller (2001) for further analysis and explanation of how this viewpoint
arises, upon applying Lagrangian averaging. We denote
$\mathbf{u}(\mathbf{x}^{\,\prime},t)=\mathbf{u}^{\,\prime}$ and begin our
investigation of the correlation dynamics by computing the ingredients of
the partial time derivative $\partial_t (v_i u^{\,\prime}_j)$,
%
\begin{eqnarray*} 
&&\hspace{-1cm}
\partial_t v_i
+
\partial_k\big(\,v_i u^k + p\delta_i^k \,
-\,\alpha^2\,u_{i,m}u^{m,k}\,\big)
=
\nu\Delta{v}_i
\,,\\&&\hspace{-1cm}
\partial_t {u^{\,\prime}}_j
+
\partial_k^{\,\prime}\big(\,{u^{\,\prime}}_j {u^{\,\prime}}^k 
+ \tilde{p}^{\,\prime}\,\delta_j^k \,\big)
+\,\alpha^2\,\partial_k\, \Big(g*{\tau^{\,\prime}}_j^k\Big)
=
\nu\Delta^{\,\prime}{u}^{\,\prime}_j
\,.
\end{eqnarray*} 
%
We cross multiply and add these equations, average the result
$\overline{(\,\cdot\,)}$ and use statistical homogeneity in the following
form, with
$\xi\equiv\mathbf{x}^{\,\prime}-\mathbf{x}$, in the traditional
KH notation, 
%
\begin{eqnarray*} 
\frac{\partial}{\partial \xi^k}\,\overline{(\,\cdot\,)}
=
\frac{\partial}{\partial {x^{\,\prime}}^k}\,\overline{(\,\cdot\,)}
=
-\,\frac{\partial}{\partial x^k}\,\overline{(\,\cdot\,)}
\,,
\end{eqnarray*} 
%
to find 
%
\begin{eqnarray*} 
\partial_t \,\overline{(\,v_i{u^{\,\prime}}_j\,)}
&-&
\frac{\partial}{\partial \xi^k}\,
\overline{\big(\,(v_i u^k 
\, -\,\alpha^2\,u_{i,m}u^{m,k}){u^{\,\prime}}_j\,\big)}
+
\frac{\partial}{\partial \xi^k}\,
\big(\,
\overline{(\,v_i\,\tilde{p}^{\,\prime})}\,\delta^k_j
-
\overline{(\,u^{\,\prime}_j\,{p})}\,\delta^k_i
\,\big)
\\
&+&
\frac{\partial}{\partial \xi^k}\,
\overline{\big(\,v_i \, ( {u^{\,\prime}}_j {u^{\,\prime}}^k 
\, +\,\alpha^2\,g*{\tau'}_j^k\,)\,\big)}
=
2\nu\Delta_\xi\overline{\big(\,(v_i {u^{\,\prime}}_j\big)}
\,,
\end{eqnarray*} 
%
where $\Delta_\xi$ is the Laplacian operator in the separation coordinate
$\xi$. Next, we symmetrize in $i,j$ and use the relation obtained from
homogeneity,
%
\begin{eqnarray*} 
\overline{\big(\,v_i {u^{\,\prime}}_j {u^{\,\prime}}^k  
+ v_j {u^{\,\prime}}_i {u^{\,\prime}}^k \,\big)}
=
-\,\overline{\big(\,{v^{\,\prime}}_i u_j u^k  
+ {v^{\,\prime}}_j u_i u^k \,\big)}
\,,
\end{eqnarray*} 
%
to find the homogeneous correlation dynamics for LANS$-\alpha$,
%
\begin{eqnarray} \label{raw-tensor-eqn}
\partial_t \,\overline{(\,v_i{u^{\,\prime}}_j+v_j{u^{\,\prime}}_i\,)}
-
\frac{\partial}{\partial \xi^k}\,
\Big(
{\cal T}_{ij}^k
-
\alpha^2\,
{\cal S}_{ij}^k
-
\Pi_{ij}^k
\Big)
=
2\nu\Delta_\xi\overline{
\big(\,v_i {u^{\,\prime}}_j+v_j {u^{\,\prime}}_i\big)}
\,.\quad
\end{eqnarray} 
%
In this equation, the symmetric tensors
${\cal T}_{ij}^k$, $\Pi_{ij}^k$ and ${\cal S}_{ij}^k$ are defined as
%
\begin{eqnarray*} 
\Pi_{ij}^k
&\equiv&
\overline{(\,v_i\,\tilde{p}^{\,\prime})}\,\delta^k_j
+
\overline{(\,v_j\,\tilde{p}^{\,\prime})}\,\delta^k_i
-
\overline{(\,u^{\,\prime}_j\,{p})}\,\delta^k_i
-
\overline{(\,u^{\,\prime}_i\,{p})}\,\delta^k_j
\,,\\
{\cal T}_{ij}^k
&\equiv&
\overline{\big( \,v_i {u^{\,\prime}}_j  + v_j {u^{\,\prime}}_i 
+
{v^{\,\prime}}_i u_j  + {v^{\,\prime}}_j u_i  
\,\big)\,u^k }
\,,\\
{\cal S}_{ij}^k
&\equiv&
\overline{
\big(\,u_{i,m}{u^{\,\prime}}_j + u_{j,m}{u^{\,\prime}}_i\,\big)u^{m,k}}
+ 
\overline{
\big(\,v_i\,g*{\tau^{\,\prime}}_j^k + v_j\,g*{\tau^{\,\prime}}_i^k\,\big)}
\,.
\end{eqnarray*} 
%

\subsection*{Imposing isotropy}

We now suppose the LANS$-\alpha$ solution is isotropic and follow the
approach of K\'arm\'an \& Howarth (1938), as refined by Robertson (1940)
and Chandrasekhar (1951) using the invariant theory of isotropic tensors.
By a standard argument, isotropy implies we may drop the
pressure-velocity tensor $\Pi_{ij}^k$. Hence, we rewrite equation
(\ref{raw-tensor-eqn}) as 
%
\begin{equation} \label{new-tensor-eqn}
\frac{\partial}{\partial t} \,{\cal Q}_{ij}
=
\frac{\partial}{\partial \xi^k}\,
\Big(
{\cal T}_{ij}^k
-
\alpha^2\,
{\cal S}_{ij}^k
\Big)
+
2\nu\Delta_\xi {\cal Q}_{ij}
\,,
\end{equation} 
%
with the corresponding definition, 
${\cal Q}_{ij} \equiv
\overline{(\,v_i{u^{\,\prime}}_j+v_j{u^{\,\prime}}_i\,)}$.  According to
their definitions, the tensors 
${\cal Q}_{ij}$ and ${\cal T}_{ij}^k$
are both symmetric and divergence-free in their indices $i,j$ for constant
$\alpha$. For consistency with the isotropy assumption, equation
(\ref{new-tensor-eqn}) implies that
${\cal S}_{ij}^k$ must {\it also} be symmetric and divergence-free in
its indices $i,j$. 

\noindent{\bf Remark.} Thus, DNS could check whether $\alpha\ne0$ is
consistent with isotropy in a given flow regime by checking whether the
divergences of ${\cal S}_{ij}^k$ vanish in its indices $i,j$.

According to the classical theory of invariants discussed in Robertson
(1940) and Chandrasekhar (1951), these three symmetric, divergence-free,
isotropic tensors may each be expressed in terms of a single defining
function. In particular, the isotropic $u-v$ autocorrelation tensor is
given by
%
\begin{eqnarray*} 
{\cal Q}_{ij}
=
{\rm curl}\,\big(Q\,\varepsilon_{ij\ell\,}\xi^\ell\,\big)
=
rQ^{\,\prime}\Big(\frac{\xi_i\xi_j}{r^2}-\delta_{ij}\Big)
-
2Q\delta_{ij}
\,,
\end{eqnarray*} 
%
with defining function $Q(r,t)$ and $Q^{\,\prime}=\partial{Q}/\partial{r}$
in the KH notation. The isotropic triple correlation tensor is
%
\begin{eqnarray*} 
{\cal T}_{ij}^k
&=&
{\rm curl}\,\Big(T\,\big(\xi_i\varepsilon_{jk\ell\,}\xi^\ell\,
+
\xi_j\varepsilon_{ik\ell\,}\xi^\ell\,\big)\Big)
\\
&=&
\frac{2}{r}T^{\,\prime}\xi_i\xi_j\xi_k
-
(rT^{\,\prime}+3T)
\Big(\xi_i\delta_{jk}+\xi_j\delta_{ik}\Big)
+
2T\delta_{ij}\xi_k
\,,
\end{eqnarray*} 
%
with defining function $T(r,t)$ and antisymmetric tensor
$\varepsilon_{ij\ell\,}$. Hence, we compute the divergence,
%
\begin{eqnarray*} 
\frac{\partial}{\partial \xi^k}\,
{\cal T}_{ij}^k
=
{\rm curl}\,\Big((rT^{\,\prime}+5T)\varepsilon_{ij\ell\,}\xi^\ell\,\Big)
=
{\rm curl}\,\Big(\frac{1}{r^4}(r^5T)^{\,\prime}\varepsilon_{ij\ell\,}\xi^\ell\,\Big)
\,.
\end{eqnarray*} 
%
This is formula (45) in Chandrasekhar (1951) and, of course, it agrees with
the corresponding formula (4.13) in Robertson (1940). Likewise, the
isotropic mean sub$-\alpha$ scale stress tensor ${\cal S}_{ij}^k$ must
also take the same form,
%
\begin{eqnarray*} 
\frac{\partial}{\partial \xi^k}\,
{\cal S}_{ij}^k
=
{\rm curl}\,\Big((rS^{\,\prime}+5S)\varepsilon_{ij\ell\,}\xi^\ell\,\Big)
=
{\rm curl}\,\Big(\frac{1}{r^4}(r^5S)^{\,\prime}\varepsilon_{ij\ell\,}\xi^\ell\,\Big)
\,,
\end{eqnarray*} 
%
with defining function $S(r,t)$. 

\noindent{\bf Remark.} Because to the presence of curl in their
definitions, the defining functions $T$ and $\alpha^2 S$ have dimensions
of energy dissipation rate, $(\overline{u^2})^{3/2}/r$. Their relations to
the defining functions in K\'arm\'an \& Howarth (1938) will be discussed
in a later section.

\subsection*{The equations governing the defining scalars}

According to Robertson (1940), the scalar defining the Laplacian of a
second order isotropic tensor is obtained by operating with 
\begin{eqnarray*} 
D
=
\Big(\frac{\partial^2}{\partial r^2}
+
\frac{4}{r}\frac{\partial}{\partial r}\Big)
=
\frac{1}{r^4}\frac{\partial}{\partial r}r^4\frac{\partial}{\partial r}
\,,
\end{eqnarray*} 
on the scalar defining the original tensor. That is,
\begin{equation}\label{DQ-rel}
D(Q)=r^{-4}(r^4Q^{\,\prime})^{\,\prime}
\,.
\nonumber
\end{equation}
The scalars defining the various second order tensors in equation
(\ref{new-tensor-eqn}) are, therefore,
\begin{eqnarray*} 
\frac{\partial Q}{\partial t}
\,,\quad
\Big(r\frac{\partial}{\partial r}+5\Big)T
=
\frac{1}{r^4}(r^5T)^{\,\prime}
\,,\quad
\Big(r\frac{\partial}{\partial r}+5\Big)S
=
\frac{1}{r^4}(r^5S)^{\,\prime}
\,,\quad
D(Q)
\,.
\end{eqnarray*} 
As Robertson (1940) points out, upon assuming isotropy, the tensor
equation (\ref{new-tensor-eqn}) is entirely equivalent to the
corresponding scalar equation. This observation proves the following:

\begin{theorem}[KH theorem for the LANS$-\alpha$ model]
\label{KH-alpha-theorem}
Let the LANS$-\alpha$ model flow be isotropic. Then, the scalar equation
%
\begin{equation} \label{KH-alpha-scalar-eqn}
\frac{\partial Q}{\partial t} 
=
\Big(r\frac{\partial}{\partial r}+5\Big)(T - \alpha^2 S)
+
2\nu  D(Q)
\,,
\end{equation} 
%
defines an exact relation among the 2nd and 3rd correlation functions, and
the stress tensor ${\cal S}_{ij}^k$.
\end{theorem}

\noindent{\bf Remark.} Formula (\ref{KH-alpha-scalar-eqn}) is the analog
for the LANS$-\alpha$ fluid equations of the KH theorem for NS turbulence.
\bigskip

Upon setting $\partial Q/\partial t = - 2\,\overline{\varepsilon_\alpha}/3$
for the energy dissipation rate in three dimensions and dropping
the viscous terms in equation (\ref{KH-alpha-scalar-eqn}) -- as
is appropriate for separation $r$ in the inertial range -- one finds the
energy balance relation for the LANS$-\alpha$ model,
\begin{equation} \label{KH-alpha-balance-eqn}
-\,\frac{2}{3}\,\overline{\varepsilon_\alpha}
=
\frac{1}{r^4}\frac{\partial}{\partial r}\Big(r^5(T - \alpha^2 S)\Big)
\,.
\end{equation} 
Here, $\overline{\varepsilon_\alpha}$ denotes the average dissipation rate
of the total kinetic energy for the LANS$-\alpha$ model, given by
$E_\alpha=\frac{1}{2}\int \mathbf{u}\cdot\mathbf{v}\, d^3x$. Integration of
the energy balance relation (\ref{KH-alpha-balance-eqn}) then proves the
following:

\begin{corollary}[``--2/15 law'' for the LANS$-\alpha$ model]
\label{-2/15 s-corollary}
In the inertial range and for arbitrary ratio $\alpha/r$,
the LANS$-\alpha$ model satisfies the ``--2/15 law,''
\begin{equation} \label{alpha-2/15-law}
-\,\frac{2}{15}\,\overline{\varepsilon_\alpha}
=
T - \alpha^2 S
\,.
\end{equation} 
\end{corollary}

\noindent{\bf Remark.} Formula (\ref{alpha-2/15-law}) is the analog for
the LANS$-\alpha$ fluid equations of Kolmogorov's ``--4/5 law'' for NS
turbulence.

\subsection*{Comparisons with NS turbulence theory for $\alpha/r\to0$}

Foias, Holm and Titi (2001b) show that solutions of the LANS$-\alpha$ model
converge to solutions of the NS equations as $\alpha\to0$ uniformly for
any positive viscosity. Therefore, to compare the KH$-\alpha$ theorem with
the results of K\'arm\'an \& Howarth (1938) for the NS equations, we may
consider the limit as $\alpha/r\to0$. In this limit, one may neglect
$\alpha^2 S$ in equations (\ref{KH-alpha-balance-eqn}) and
(\ref{alpha-2/15-law}).  We follow Robertson (1940) in identifying the KH
double correlation scalars
$f(r,t)$, $g(r,t)$ as
\begin{equation} \label{KH-alpha-double}
(\overline{u^2})f = Q
\,,\quad\hbox{and}\quad
(\overline{u^2})g = \frac{1}{2}rQ^{\,\prime}
\,.
\end{equation} 
Likewise, the KH triple correlation scalars $h$, $k$ and $q$ are
identified in terms of $T(r,t)$, as $(\overline{u^2})^{3/2}h = rT/2$ for
$h$, as well as,
\begin{equation} \label{KH-alpha-triple}
(\overline{u^2})^{3/2}k = -\,\frac{1}{r^4}\int_0^r s^4 T\,ds
\quad\hbox{and}\quad
(\overline{u^2})^{3/2}q 
= 
\frac{1}{8r^4}\int_0^r s^4 T\,ds\, -\, \frac{r}{4}\,T
\,,\nonumber
\end{equation} 
for $k$ and $q$. Using the relation $T=2(\overline{u^2})^{3/2}h/r$ yields,
\begin{equation} \label{KH-alpha-hT-rel}
\Big(r\frac{\partial}{\partial r}+5\Big)\,T
=
2(\overline{u^2})^{3/2}
\Big(\frac{\partial}{\partial r}+\frac{4}{r}\Big)\,h
\,.
\end{equation} 
Hence, when $\alpha/r\to0$, equation (\ref{KH-alpha-scalar-eqn}) of the
KH$-\alpha$ theorem \ref{KH-alpha-theorem} yields equation (51) of
K\'arm\'an
\& Howarth (1938), namely, the KH equation,
\begin{equation} \label{KH-eqn51}
(\overline{u^2})\,\frac{\partial f}{\partial t} 
=
\Big(\frac{\partial}{\partial r}+\frac{4}{r}\Big)
\bigg[ 
2(\overline{u^2})^{3/2}\,h
+
2\nu\, (\,\overline{u^2}\,) \,
\frac{\partial f}{\partial r}
\bigg]
\,.
\end{equation} 
One may refer to Monin \& Yaglom (1975) page 122, for the KH equation in
their notation. When the factor 1/6 relating third order autocorrelation
functions and structure functions is introduced, this is also equation (3)
of Kolmogorov (1941b), leading to the Kolmogorov's $-4/5$ law for
Navier-Stokes fluids. See also Landau \& Lifschitz (1987) for additional
discussion of this fundamental result.

Thus, the limit $\alpha/r\to0$ of formula  (\ref{KH-alpha-scalar-eqn}) of
the KH$-\alpha$ theorem recovers the expected classical results for
homogeneous, isotropic, NS turbulence.

\section{Conclusions}

\noindent{\bf Main results.} 
The main results of the paper are, as follows. 
\begin{description}

\item$\bullet\quad$
Equation (\ref{raw-tensor-eqn}) for the LANS$-\alpha$ dynamics of the
velocity autocorrelation functions with homogeneous statistics shows that
the alpha model's preservation of transport structure
extends to preserving the form of these transport equations, as well,
modulo a shift in the term involving third moments that
accommodates the contribution from the sub-alpha scale stress tensor. 

\item$\bullet\quad$
The modifications of the KH theorem and Kolmogorov's energy
balance law for NS to include the effects of alpha-filtering in the
LANS$-\alpha$ model are relatively simple and straightforward.

\begin{description}
\vspace{-2mm}\item$\bullet\quad$
Equation (\ref{KH-alpha-scalar-eqn}) in the KH$-\alpha$ theorem gives the
alpha-analog of the Karman-Howarth theorem for isotropic turbulence. 
\item$\bullet\quad$
As a corollary of the KH$-\alpha$ theorem, equation (\ref{alpha-2/15-law})
gives the ``$-2/15$ law'' alpha-analog of Kolmogorov's ``$-4/5$ law'' for
energy balance in the inertial range. (The relative 1/6 in these
coefficients arises from the 1/6 relation between autocorrelation
functions and structure functions in isotropic turbulence.)
\item$\bullet\quad$
The second term in the $-2/15$ law in equation (\ref{alpha-2/15-law}), the
${\cal S}-$term which is proportional to $\alpha^2$, is very reminiscent
of the quantity that appears in the corresponding $-2$ law for enstrophy
cascade in 2D turbulence. The latter expression contains two powers of
vorticity and one power of velocity. For example, see the Appendix B of
Eyink (1996), where the identity is derived in detail. In the same way,
the ${\cal S}-$term in (\ref{KH-alpha-scalar-eqn}) contains two powers of
velocity gradient and one of velocity. It should thus be the dominant
term (compared to the first ${\cal T}-$term) when $r<\alpha$ and the
velocity is nearly smooth. It then corresponds to an``enstrophy-like"
cascade, in agreement with the considerations of Foias, Holm and Titi
(2001a) that determine the $k^{-3}$  spectrum in that range.%
\footnote{I am grateful to G. Eyink for this observation.}

\end{description}\end{description}

These exact results confirm that alpha-filtering affects the homogeneous
isotropic statistical properties of NS only at separations $r$ of order
$O(\alpha)$, or smaller. This is because of the exponential decrease with
separation in the Green's function for the Helmholtz operator. The same
exponential decrease with separation applies for the dynamics of the
higher order autocorrelation functions and structure functions, whose
anomalous scaling properties for NS indicate the intermittency of
turbulence.  Consequently, one may expect the intermittency of the alpha
model will be the {\it same} as for NS at large separations, for which
$r/\alpha\gg1$. This conclusion may apply more generally for other types
of filtering, as well.

Some aspects of these results were not entirely unexpected. After all, 
one could enforce homogeneity and isotropy on the velocity
correlation functions for {\it any} fluid model (e.g., an LES model, or the
2$^{nd}$ grade fluid model). One could then compute its corresponding KH
equation by the same classical  methods as those used here. In general,
this procedure could introduce complications, such as requiring several
more scalar defining functions. Remarkably, the KH$-\alpha$ theorem for the
case of constant alpha may be written at the cost of just one additional
scalar function, S(r,t), arising from the sub-alpha scale stresses. The
LANS$-\alpha$ model preserves the fundamental transport structure of the
NS equations. It also preserves the form (\ref{new-tensor-eqn}) of the
transport relations for the velocity autocorrelation functions. Thus, the
interpretations of its consequences turned out to be relatively
straightforward in this case. 

\noindent{\bf Helical turbulence.} These results readily
extend to helical turbulence, and are expected also to extend to any
desired higher order autocorrelation equations. For example,
one may expect that the fundamental relations for the third moments in
helical turbulence obtained in Chkhetiani (1996) will also be expressed
simply and concisely in the LANS$-\alpha$ model formulation. This will be
discussed elsewhere.

\noindent{\bf Spectral representation.} The spectral representation of
these correlation dynamics is also worth examining further,
especially because the inverse of the Helmholtz operator is algebraic in
spectral space. As mentioned earlier, this direction was partially
developed in Foias, Holm and Titi (2001a), where the $k^{-5/3}\to k^{-3}$
roll-off in the LANS$-\alpha$ energy spectrum was discovered in the
spectral range for which $k\alpha<1$ passes to $k\alpha>1$. However, much
remains to be studied about how the alpha-filtering in the LANS$-\alpha$
model affects (or preserves) other fundamental aspects of turbulence
modeling, while at the same time making the resulting regularized
turbulence model more computationally accessible.

\noindent{\bf Loitsyanski invariant.} On the basis of our earlier
reasoning that the LANS$-\alpha$ model results should recover the NS
results as $\alpha/r\to0$, we expect that the LANS$-\alpha$ predictions
should agree with NS theory at large separation, particularly in regard to
the behavior of the Loitsyanki integral.

\noindent{\bf Future directions.}%
\footnote{We are grateful to U. Frisch and M. Vergassola for their 
helpful suggestions here.}
The KH$-\alpha$ theorem confirms analytically that alpha-filtering leaves
the velocity autocorrelation statistics invariant at sufficiently large
separations $\alpha/r\ll1$  (and, thus, could be expected to preserve
intermittency at those separations) for homogeneous isotropic turbulence.
This has three promising implications for future research directions:
 \vspace{-2mm}
\begin{description}
\item
(1) The alpha $\to\infty$ limit equation could be used to examine the 
small-separation effects of alpha on the statistics.

\item
(2) One may study the dependence on the order of the correlation function
of the transition from the $\alpha$-dominated to the $\alpha$-irrelevant
regime. This would determine the $\alpha$-dependence of the scaling
regions.

\item
(3) Needing only to compute scales significantly larger than a fixed value
of alpha in LANS$-\alpha$ should provide more dynamic range for DNS 
than in the NS case, when alpha vanishes. This suggests one might fix
alpha in the LANS$-\alpha$ model and use DNS to seek evidence of
{\it universality at large scales}, especially in turbulence decay. 

\end{description}
\bigskip

\noindent{\bf Acknowledgments.}  
We are grateful to S. Chen, G. Eyink, U. Frisch, R. M. Kerr, J. E. Marsden,
S. Shkoller, M. Vergassola and B. Wingate for their constructive comments
and encouragement. This work was supported by DOE, under contract
W-7405-ENG-36.


\section*{References}

\begin{description}

\item
{\it Analogy between predictions of Kolmogorov and Yaglom},\\
R. A. Antonia, M. Ould-Rouis, F. Anselmet \& Y. Zhu,
{\it J. Fluid Mech.} {\bf 332} (1997) 395-409.

\item
{\it Correlation functions in isotropic and anisotropic turbulence: The
role of the symmetry group},\\
I. Arad, V. S. L'vov and I. Procaccia,
{\it Phys. Rev. E} {\bf 59} (1999) 6753-6765.


\item
{\it An integrable shallow water equation with peaked solitons},\\
R. Camassa and D.D. Holm, Phys. Rev. Lett. {\bf71} (1993) 1661-1664.

\item 
{\it The invariant theory of isotropic turbulence in
magneto-hydrodynamics},\\
S. Chandrasekhar, 
{\it Proc. Roy. Soc. London A} {\bf 204} (1951) 435-449.

\item
{\it The Camassa-Holm equations as a closure model for
turbulent channel and pipe flows},\\
S. Y. Chen, C. Foias, D. D. Holm, E. J. Olson, E. S. Titi, \&
S. Wynne,
{\it Phys. Rev. Lett.} {\bf 81} (1998) 5338-5341.

\item
{\it The Camassa-Holm equations and turbulence in pipes and channels},\\
S. Y. Chen, C. Foias, D. D. Holm, E. J. Olson, E. S. Titi, \&
S. Wynne,
{\it Physica D} {\bf133} (1999a) 49-65.

\item
{\it A connection between the Camassa-Holm equations and turbulence
in pipes and channels},\\
S. Y. Chen, C. Foias, D. D. Holm, E. J. Olson, E. S. Titi, \&
S. Wynne, 
{\it Phys. Fluids} {\bf11} (1999b) 2343-2353.

\item
{\it Direct numerical simulations of the Navier-Stokes alpha model},\\
S. Y. Chen, D. D. Holm, L. G. Margolin \& R. Zhang,
{\it Physica D} {\bf133} (1999c) 66-83.

\item 
{\it On the third moments in helical turbulence},\\
O. G. Chkhetiani, {\it JETP Lett.} {\bf 63} (1996) 808-812.

\item 
{\it Thermodynamics, stability, and boundedness of fluids of
complexity-2 and fluids of second grade},\\ 
J. E. Dunn \& R. L. Fosdick,
{\it Arch. Rat. Mech. \& Anal.} {\bf 56}
(1974) 191-252.

\item
{\it Exact results on stationary turbulence in 2D: Consequences of
vorticity conservation}\\ 
G. Eyink,
{\it Physica D} {\bf 91} 97-142 (1996).

\item
{\it The Navier-Stokes-alpha model of fluid turbulence},\\
C. Foias, D. D. Holm and E. S. Titi,
{\it Physica D}, to appear (2001a).

\item
{\it The Three Dimensional Viscous Camassa--Holm Equations,
and Their Relation to the Navier--Stokes Equations and
Turbulence Theory},\\
C. Foias, D. D. Holm and E. S. Titi,
{\it J. Dyn. and Diff. Eqns.}, to appear (2001b).

\item
{\it Turbulence: The Legacy of A. N. Kolmogorov},\\ 
U. Frisch, (Cambridge University Press, 1995).

\item
{\it  Longitudinal structure functions in decaying and forced turbulence},\\
D. Fukayama, T. Oyamada, T. Nakano, T. Gotoh and K. Yamamoto,
{\it Journal of the Physical Society of Japan} {\bf 69} (2000) 701-715.

\item
{\it Exact relationship for third-order structure functions in helical
flows},\\ 
T. Gomez, H. Politano \& A. Pouquet,
{\it Phys. Rev. E} {\bf 61} (2000) 5321-5325.

\item
{\it Applicability of Kolmogorov's and Monin's equations of turbulence},\\
R. J. Hill \& O. N. Boratav,
{\it J. Fluid Mech.} {\bf 353} (1997) 67-81.

\item
{\it Next-order structure functions},\\
R. J. Hill \& O. N. Boratav,
{\it Phys. Fluids} {\bf 13} (2001) 276-283.

\item
{\it Fluctuation effects on 3D Lagrangian mean
and Eulerian mean fluid motion},\\ 
D.D. Holm, {\it Physica D}, {\bf133} (1999) 215-269.

\item
{\it Averaged Lagrangians and the mean dynamical effects of
fluctuations in continuum mechanics},\\
D.D. Holm, {\it Physica D}, submitted,\\
http://xxx.lanl.gov/abs/nlin.CD/0103035.

\item
{\it Euler--Poincar\'e models of ideal fluids
with nonlinear dispersion},\\
D.D. Holm, J.E. Marsden and T.S. Ratiu,
Phys. Rev. Lett. {\bf 80} (1998a) 4173-4177.

\item
{\it The Euler--Poincar\'e equations and semidirect products
with applications to continuum theories},\\
D.D. Holm, J.E. Marsden, T.S. Ratiu,
Adv. in Math. {\bf 137} (1998b) 1-81.

\item
{\it On the statistical theory of isotropic turbulence},\\
T. von K\'arm\'an \& L. Howarth 
{\it Proc. Roy. Soc. London A} {\bf 164} (1938) 192-215.

\item
{\it The local structure of turbulence in incompressible viscous fluid
for very large Reynolds number},\\ 
A. N. Kolmogorov, {\it Dokl. Akad. Nauk SSSR} {\bf 30} (1941a) 9-13
(reprinted in {\it Proc. Roy. Soc. London A} {\bf 434} (1991) 9-13).

\item
{\it Dissipation of energy in locally isotropic turbulence},\\ 
A. N. Kolmogorov, {\it Dokl. Akad. Nauk SSSR} {\bf 32} (1941b) 16-18
(reprinted in {\it Proc. Roy. Soc. London A} {\bf 434} (1991) 15-17).

\item
{\it A refinement of previous hypotheses concerning the local structure of
turbulence in a viscous incompressible fluid at high Reynolds number},\\ 
 A. N. Kolmogorov, {\it J. Fluid Mech.} {\bf 13} (1962) 82-85.

\item
{\it Fluid Mechanics},\\ L. D. Landau \& E. M. Lifschitz (Pergamon Press: New
York, 2nd Edition 1987).

\item
{\it A note on Kolmogorov's third-order structure-function law, the
local isotropy hypothesis and the pressure-velocity correlation},\\ 
E. Lindborg {\it J. Fluid Mech.} {\bf 326} (1996) 343-356.

\item
{\it Introduction to Mechanics and Symmetry}, 2nd Edition,\\ 
J.E. Marsden and T.S. Ratiu, (Springer, New York, 1999).

\item
{\it Global well-posedness for the Lagrangian averaged
Navier-Stokes (LANS$-\alpha$) equations on bounded domains},\\
J. E. Marsden and S. Shkoller,
{\it Proc. Roy. Soc. London A}, to appear.

\item
{\it Numerical simulations of homogeneous turbulence
using the Lagrangian averaged Navier-Stokes equations},\\
K. Mohseni, S. Shkoller, B. Kosovi\'c, J. E. Marsden, D. Carati, A.
Wray, and R. Rogallo, 
Center for Turbulence Research, Proceedings of the Summer Program, 2000,
271-283.

\item
{\it Statistical Fluid Mechanics: Mechanics of Turbulence},\\
A. S. Monin \& A. M. Yaglom, (MIT Press, Cambridge, 1975).

\item
{\it von K\'arm\'an-Howarth equation for magnetohydrodynamics and its
consequences on third-order longitudinal structure and correlation
functions},\\ 
H. Politano \& A. Pouquet, {\it Phys. Rev. E} {\bf 57} (1998) R21-R25.

\item
{\it The invariant theory of isotropic turbulence},\\
H. P. Robertson, {\it Proc. Camb. Phil. Soc.} {\bf 36} (1940) 209-223.

\end{description}

\end{document}